\documentclass[twocolumn,superscriptaddress, showpacs] {revtex4}
\usepackage{graphicx}
\usepackage{amsmath}
\usepackage{dsfont}
\usepackage{amssymb}

\begin{document}

\title[Short Title]{One-step achievement of robust multipartite Greenberger-Horne-Zeilinger state and controlled-phase gate via Rydberg interaction}
\author{Xiao-Qiang Shao\footnote{E-mail: xqshao@yahoo.com}}
\affiliation{School of Physics, Northeast Normal University
Changchun 130024, People's Republic of China}
\affiliation{Centre for Quantum Technologies, National University of Singapore, 3 Science Drive 2, Singapore 117543}
\author{Tai-Yu Zheng}
\affiliation{School of Physics, Northeast Normal University
Changchun 130024, People's Republic of China}
\author{C. H. Oh}
\affiliation{Centre for Quantum Technologies, National University of Singapore, 3 Science Drive 2, Singapore 117543}
\author{Shou Zhang}
 \affiliation{Department of Physics, College of Science,
Yanbian University, Yanji, Jilin 133002, People's Republic of China}
\begin{abstract}
We present a proposal for generation of  a robust tripartite Greenberger-Horne-Zeilinger state among three-individual neutral Rydberg atoms. By modulating the relation between two-photon detuning and Rydberg interaction strength $U_{ij}(r)$,  an effective Raman coupling is obtained between the hyperfine ground states $|F=2,M=2\rangle$ of three $^{87}$Rb atoms and the Rydberg states $|rrr\rangle$ via the third-order perturbation theory. This method is also capable of implementing a three-qubit controlled-phase gate with each qubit encoded into the hyperfine ground states $|F=1,M=1\rangle$ and $|F=2,M=2\rangle$. As an extension, we generalize our scheme to the case of multipartite GHZ state and quantum gate in virtue of high-order perturbation theory.
\end{abstract}
 \maketitle \maketitle
\section{introduction}
In quantum physics, one of the most essential
features is termed quantum entanglement. The counterintuitive properties of entanglement in quantum theory had ever induced a dispute over the physical reality \cite{einstein}, but eventually entanglement was verified experimentally  and recognized as a valid, fundamental feature of quantum mechanics.  Einstein-Podolsky-Rosen entanglement state and Schr\"{o}dinger cat state has now become two well-known bipartite entanglements. Compared with the bipartite entanglement, the multipartite entanglement possesses more peculiar properties.
Greenberger-Horne-Zeilinger (GHZ) state, named after D. Greenberger, M. A. Horne and A. Zeilinger in 1989 \cite{ghz}, is a type of canonical multipartite entangled quantum state which involves at least three subsystems. These states are defined to be maximally entangled because they maximally violate the famous Bell-type inequalities and exhibit strong nonlocality \cite{bell}. Therefore the correlations in GHZ states are especially useful in some quantum information tasks such as multipartner quantum cryptography \cite{hillery} and communication complexity \cite{wangc,xia}.

\begin{figure}
\centering\scalebox{0.14}{\includegraphics{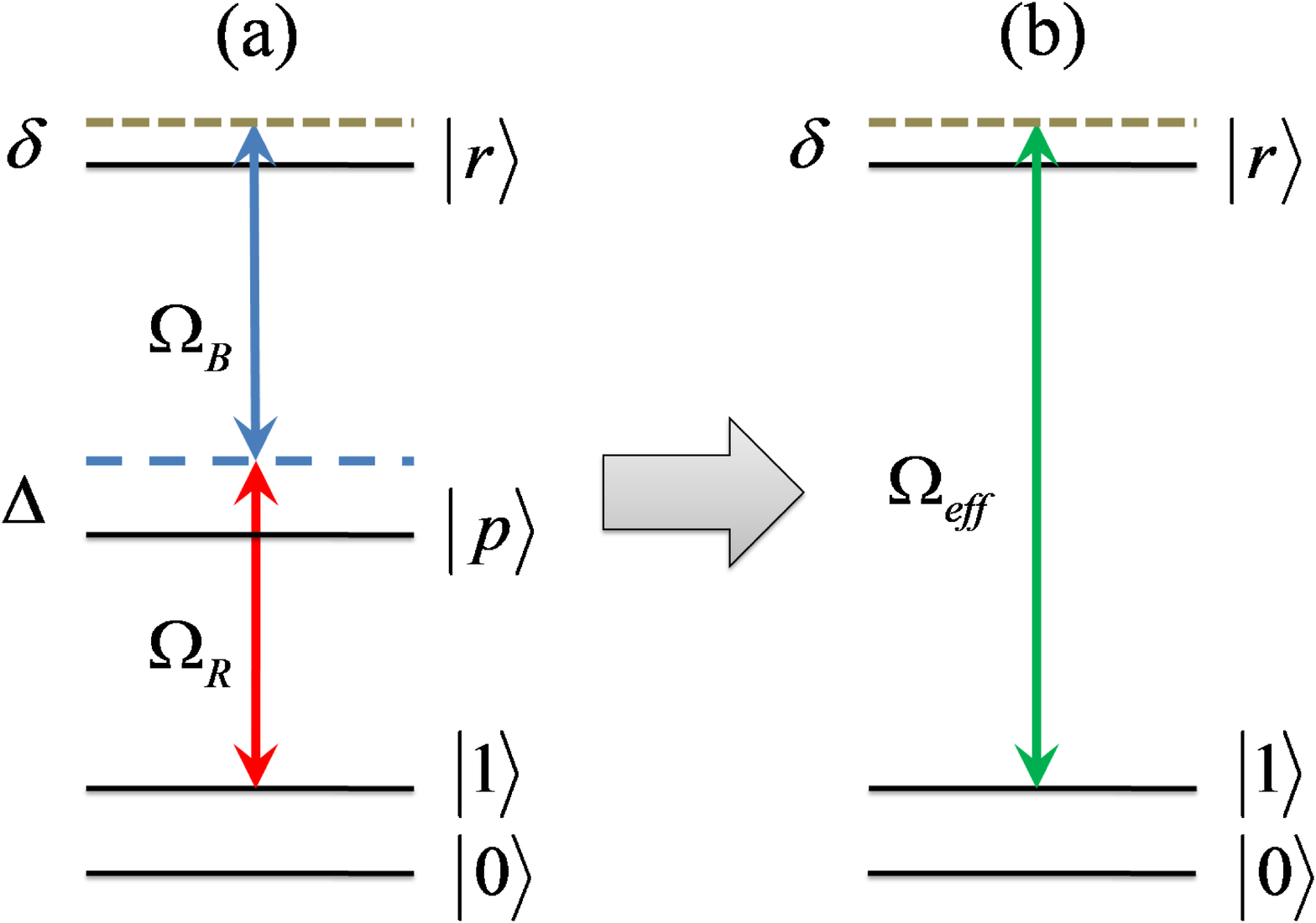} }
\caption{\label{p0}(Color online) Schematic view of atomic level configuration. (a): The
atom is driven to the Rydberg state $|r\rangle$ from ground state $|1\rangle$ via a two-photon transition, where the red laser with Rabi frequency $\Omega_R$ couples to the transition
$|1\rangle$ to the intermediate state $|p\rangle$ (which
is blue-detuned by $\Delta$) and
 the blue laser with Rabi frequency $\Omega_B$ drives the transition $|p\rangle$ to $|r\rangle$ (which is
red-detuned by $\Delta-\delta$). (b): An effective model which suppresses the spontaneous emission of optical state $|p\rangle$, given that the Rydberg state lifetime is
much longer than the optical state lifetime $\gamma_r\ll \gamma_p$, where the effective Rabi frequency $\Omega_{eff}=\Omega_R\Omega_B/\Delta.$}
\end{figure}
On the other hand, as a central and fundamental
task for actualizing a quantum computer, physical implementation of universal set of quantum gate lies at the heart of quantum computation. Although the principle of universal quantum computation allows to construct an arbitrary quantum gate with a series of single qubit
operations along with two-qubit gates \cite{divi}, this kind of composition becomes inefficient as applied to a multi-qubit gate for the procedure will become more complicated and the quantum system will be more susceptible
to the environment, since uncontrolled interaction or decoherence exists in each
gate in the experiment. Thus direct realization of a multi-qubit quantum gate can greatly simplify the procedures and enhance efficiency for quantum information processing. In particular, the multi-qubit quantum
gates usually play a key role in quantum algorithms \cite{vander} and quantum
error-correction protocols \cite{cory,saro}.

In this article, we investigate the potential application of neutral Rydberg atoms on preparation of quantum entanglement and quantum computation.  The advantage for adopting neutral atom as qubit is that the quanutm information
is encoded into the stable hyperfine ground states and distant atoms interact with each other through Rydberg blockade \cite{saff}. The Rydberg blockade regime is based on the assumption that one excited
atom causes a sufficiently large energy shifts of Rydberg
states in a neighboring atom to effectively detune it away from
resonance and fully block its excitation by a laser field. This mechanism was originally proposed by Jaksch {\it et al.} for implementing fast two-qubit quantum gates for neutral atoms \cite{jaks},
then it was observed experimentally \cite{urban} and has been explored to perform various tasks such as
entangled
state preparation \cite{wilk}, quantum algorithms \cite{chen}, quantum
simulators \cite{weimer}, and efficient quantum repeaters \cite{han}. Different from previous works with a strong Rydberg blockade \cite{ms,isen,wu,zhang}, we show here that a intermediate Rydberg interaction with a comparable magnitude to the two-photon detuning is especially functional for creation of multipartite GHZ state and controlled-phase gate.

This manuscript is structured as follows. In section II, we derive the effective Hamiltonian for preparation of tripartite GHZ state and three-qubit controlled-phase gate, and also investigate the effect of decoherence induced by spontaneous emission of Rydberg state. In section III, we generalize the scheme to realization of multipartite GHZ state and multi-qubit controlled phase gate. In section IV, we give a conclusion of this manuscript.

\section{Tripartite GHZ state and controlled-phase gate}
The system consists of three identical $^{87}$Rb atoms trapped in three separate microscopic dipole traps, and the relevant configuration of atomic level is illustrated in Fig.~\ref{p0}(a), where the quantum states $|0\rangle$ and $|1\rangle$
correspond to atomic levels $|F=1,M=1\rangle,$ and $|F=2,M=2\rangle$ of $5S_{1/2}$ manifold, and the optical state $|p\rangle=|F=2,M=2\rangle$ of $5P_{1/2}$  and Rydberg state $|r\rangle=|F=3,M=3\rangle$ of $58D_{3/2}$.
 In order to excite from $|1\rangle$ to $|r\rangle$ without population of the optical state $|p\rangle$,
 we apply a $\pi$-polarized laser, with Rabi frequency $\Omega_R$, blue-detuning $\Delta$, to drive the transition from ground state $|1\rangle$ to optical state $|p\rangle$, and a $\sigma^+$-polarized laser,  with Rabi frequency $\Omega_B$, red-detuning $\Delta-\delta$, to couple to the transition $|p\rangle$ to $|r\rangle$. The Hamiltonian of system in a rotating frame reads \cite{mur,rao}
 \begin{figure}
\scalebox{0.55}{\includegraphics{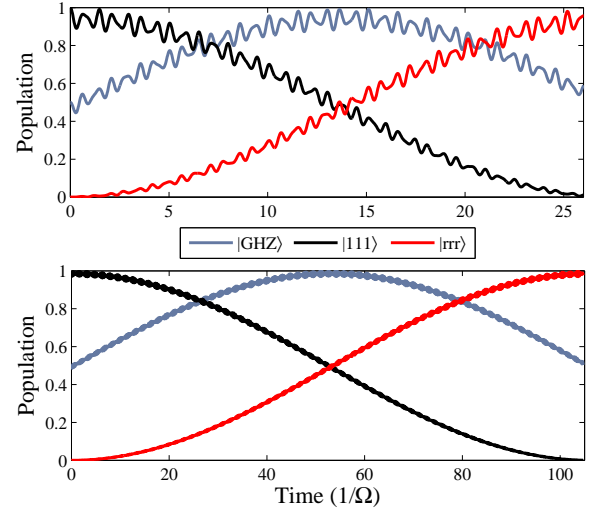} }
\caption{\label{p1}(Color online) The populations of $|111\rangle$, $|rrr\rangle$,  $|{\rm GHZ}\rangle=\frac{1}{\sqrt{2}}(|111\rangle+i|rrr\rangle)$ as a function of time in unites of $\Omega^{-1}$ for fixed $U=\delta$. The upper panel corresponds to $\Delta=10\Omega$, while the lower panel corresponds to $\delta=20\Omega$.}
\end{figure}
\begin{eqnarray}\label{111}
\hat{H}_I&=&\sum_{i=1}^3-(\Delta_i+\frac{i}{2}\gamma_p)|p\rangle_{i}\langle p|-(\delta_i+\frac{i}{2}\gamma_r)|r\rangle_{i}\langle r|
\nonumber\\
&&+\Omega^i_R(|1\rangle_{i}\langle p|+|p\rangle_{i}\langle 1|)+\Omega^i_B(|p\rangle_{i}\langle r|+|r\rangle_{i}\langle p|)\nonumber\\
&&+\sum_{i\neq j}U_{ij}(r)|rr\rangle_{ij}\langle rr|,
\end{eqnarray}
where the dipole-dipole potential
with energy $U_{ij}(r)=C_3/r^3$ is non-zero when atoms $i,j$ simultaneously occupy
the Rydberg state $|r\rangle$. $\gamma_p$ and $\gamma_r$ are the decay rates of spontaneous emission for optical state $|p\rangle$ and Rydberg state $|r\rangle$, respectively.
Since the Rydberg state lifetime is
much longer than the optical state lifetime $\gamma_r\ll \gamma_p$, we can adiabatically eliminate the state $|p\rangle$ by setting $\Delta\gg \Omega_R, \Omega_B$, then the Hamiltonian is obtained
through the second-order perturbation theory:
\begin{eqnarray}\label{222}
\hat{H}^{'}_{I}&=&\sum_{i=1}^3(-\delta_i+\frac{{\Omega^i_B}^2}
{\Delta_i}-\frac{i}{2}\gamma_r)|r\rangle_{i}\langle r|
+\frac{\Omega^i_R\Omega^i_B}{\Delta_i}(|1\rangle_{i}\langle r|\nonumber\\
&&+|r\rangle_{i}\langle 1|)+\frac{{\Omega^i_R}^2}{\Delta_i}|1\rangle_{i}\langle1|+\sum_{i\neq j}U_{ij}(r)|rr\rangle_{ij}\langle rr|.
\end{eqnarray}
The stark-shift term ${\Omega_B^i}^2/\Delta_i$ can be absorbed in $\delta_i$ or canceled
together with ${\Omega_R^i}^2/\Delta_i$ via introducing ancillary levels. Then the model reduces to a three-level system as shown in Fig.~\ref{p0}(b). In what follows, we first neglect the spontaneous emission of Rydberg state and consider a ideal situation with equal Rydberg interaction $U_{ij}(r)=U$ for each pair of atoms. Supposing the atoms are initially prepared in state $|111\rangle$, we then expand the Hamiltonian in corresponding subspace
 $\{|111\rangle,\frac{1}{\sqrt{3}}(|r11\rangle+|1r1\rangle+|11r\rangle),
 \frac{1}{\sqrt{3}}(|1rr\rangle+|r1r\rangle+|rr1\rangle)
,|rrr\rangle\}$ as

\begin{figure*}
\scalebox{0.55}{\includegraphics{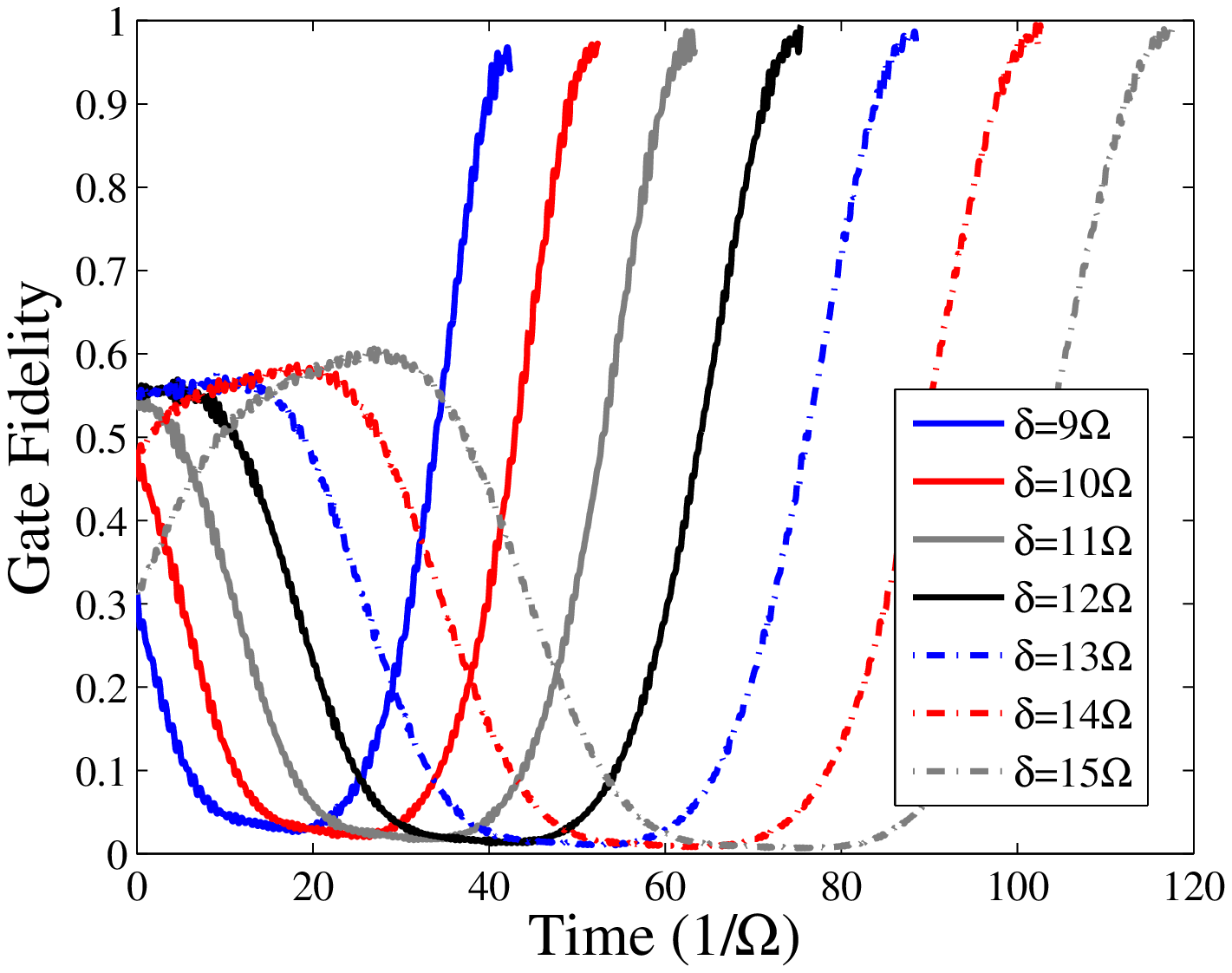} }
\scalebox{0.55}{\includegraphics{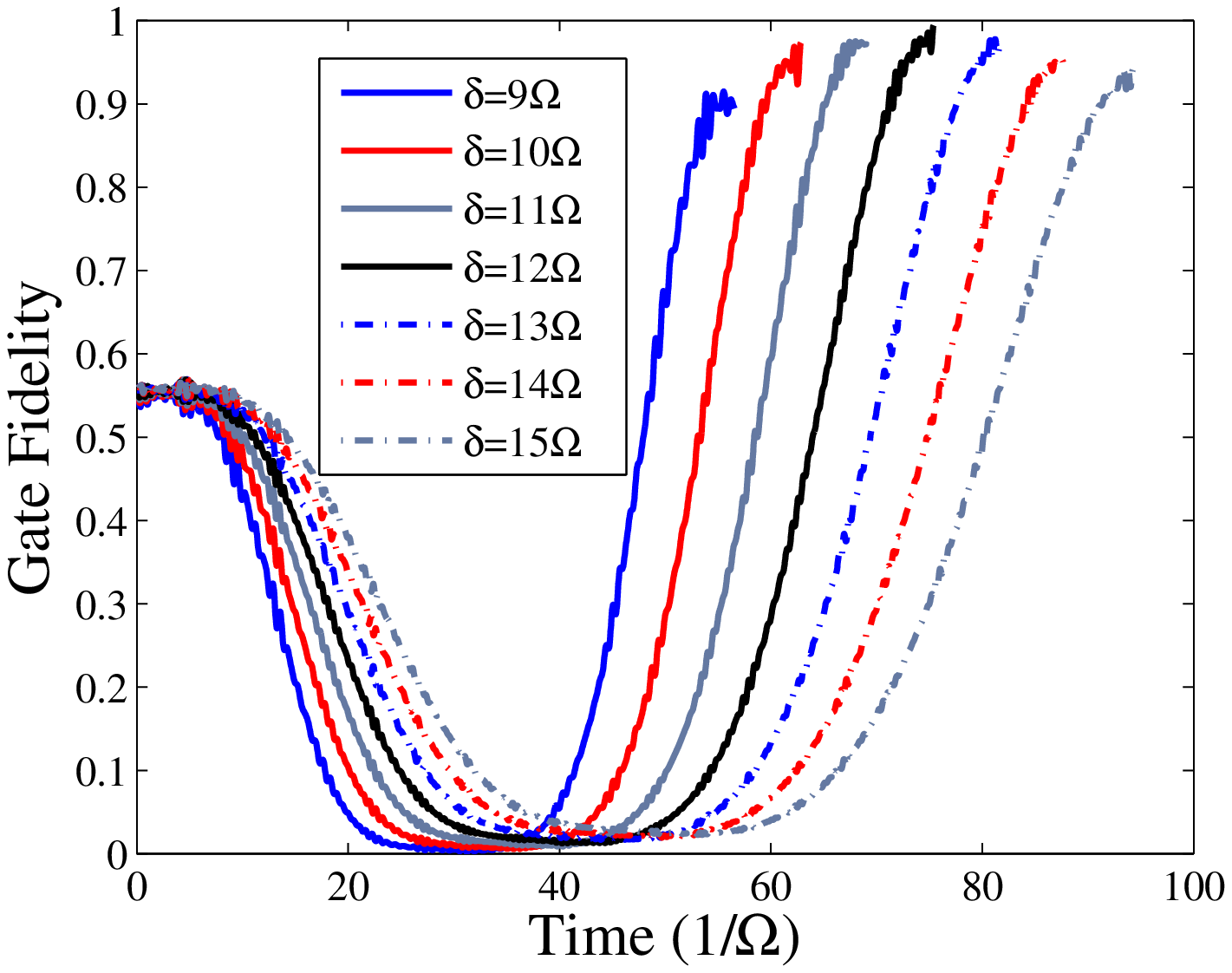} }
\caption{\label{pzz}(Color online) Gate fidelity versus interaction time. The different curves correspond to different
values of the two-photon detuning $\delta$. Left panel: Locally unitary
equivalent controlled-phase gate. Right panel: Standard controlled-phase gate. }
\end{figure*}

\begin{eqnarray}\label{one11}
\hat{H}^{\rm I}_{3}=\left[\begin{array}{c c c c}
0 & \sqrt{3}\Omega_{eff} &0 & 0 \\
\sqrt{3}\Omega_{eff} & -\delta & 2\Omega_{eff} & 0  \\
0 & 2\Omega_{eff} & U-2\delta &\sqrt{3}\Omega_{eff}  \\
0 & 0 &\sqrt{3}\Omega_{eff}  & 3U-3\delta \\
\end{array}
\right].
\end{eqnarray}
where  $\Omega_{eff}=\Omega_R^i\Omega_B^i/\Delta$.  Eq.~(\ref{one11}) describes
a climbing process from the ground state $|111\rangle$ to the three-excited state $|rrr\rangle$. If the Rydberg interaction $U$ is much stronger than the two-photon detuning
$\delta$, the state $|111\rangle$ can only couple to the collective single-excited state
$\frac{1}{\sqrt{3}}(|r11\rangle+|1r1\rangle+|11r\rangle)$, and this relates to Rydberg blockade regime, where the population of bi-excited state is prohibited. However, we here take into account another regime where the strength of Rydberg interaction has a comparable magnitude to the two-photon detuning. Reexamining Eq.~(\ref{one11}) we find the skew diagonal
is mirror symmetric, thus it is possible to transfer the quantum information between
$|111\rangle$ and $|rrr\rangle$ by adjusting the strengths of diagonal elements. In connection with current situation, we chose the parameters satisfying $U=\delta$. This selection will result in an interesting model, i.e. $|111\rangle$ and $|rrr\rangle$ couple to the collective single- and bi-excited states with the same detuning $-\delta$, while these two collective excited states resonantly interact with each other. If we further assume $\delta\gg 2\Omega_{eff}$, an effective Raman coupling between the ground state $|111\rangle$ and three-excited Rydberg  state $|rrr\rangle$ is achieved as follow
\begin{eqnarray}\label{444}
\hat{H}^{\rm I}_{3}&=&\frac{3\Omega^2_{eff}}{\delta}(|111\rangle\langle 111|+|rrr\rangle\langle rrr|)\nonumber\\&&+\frac{6\Omega^3_{eff}}{\delta^2}(|111\rangle\langle rrr|+|rrr\rangle\langle 111|),
\end{eqnarray}
where $3\Omega^2_{eff}/\delta$ originates from the stark-shift of $|111\rangle(|rrr\rangle)$ due to the dispersive interaction with collective single(bi)-excited state, and $6\Omega^3_{eff}/\delta^2$ is the effective coupling strength calculated by third-order perturbation theory. It is not difficult to see that a GHZ state $(|111\rangle+i|rrr\rangle)/\sqrt{2}$ will be generated at the appropriate time $6\Omega^3t/\delta^2=(k+1/4)\pi$, where we have shorten $\Omega_{eff}$ as $\Omega$. In Fig.~\ref{p1}, we plot the populations of corresponding states $|111\rangle$, $|rrr\rangle$,  $|{\rm GHZ}\rangle=(|111\rangle+i|rrr\rangle)/\sqrt{2}$ as a function of time in unites of $\Omega^{-1}$ for fixed $U=\delta$, where the upper panel works in the regime $\delta=10\Omega$ and the lower panel endowed a larger detuning $\delta=20\Omega$. We
see that under the given parameters the full and the effective dynamics Eqs~(\ref{222}) and (\ref{444}) of the system are in excellent agreement. Note one may also map $|{\rm GHZ}\rangle=(|111\rangle+i|rrr\rangle)/\sqrt{2}$ to $|{\rm GHZ^{'}}\rangle=(|111\rangle-|000\rangle)/\sqrt{2}$ through a similar effective Raman coupling process between $|rrr\rangle$ and $|000\rangle$ or a series of single-qubit
operations $|r\rangle_i\rightleftarrows|0\rangle_i$.

Now we apply the above mechanism to implement a three-qubit controlled-phase gate. This quantum gate is closely related to Toffoli gate which is valuable in complex quantum algorithms such as Shor's
algorithm and Grover's algorithm. The quantum information is stored in two hyperfine ground states $|0\rangle$ and $|1\rangle$ separated in frequency by 6.8 GHz. Consider a general input state
\begin{figure*}
\scalebox{0.55}{\includegraphics{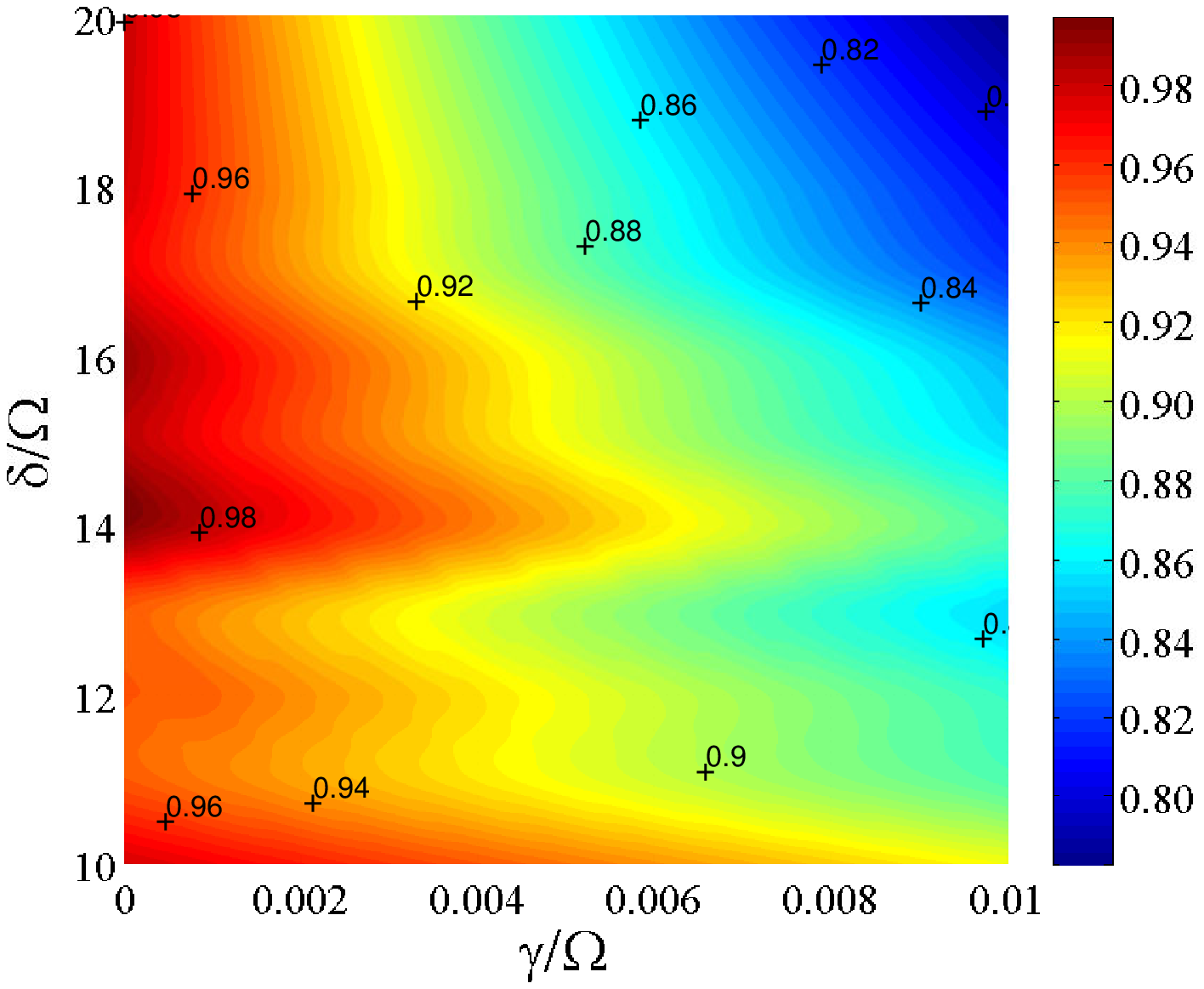} }
\scalebox{0.55}{\includegraphics{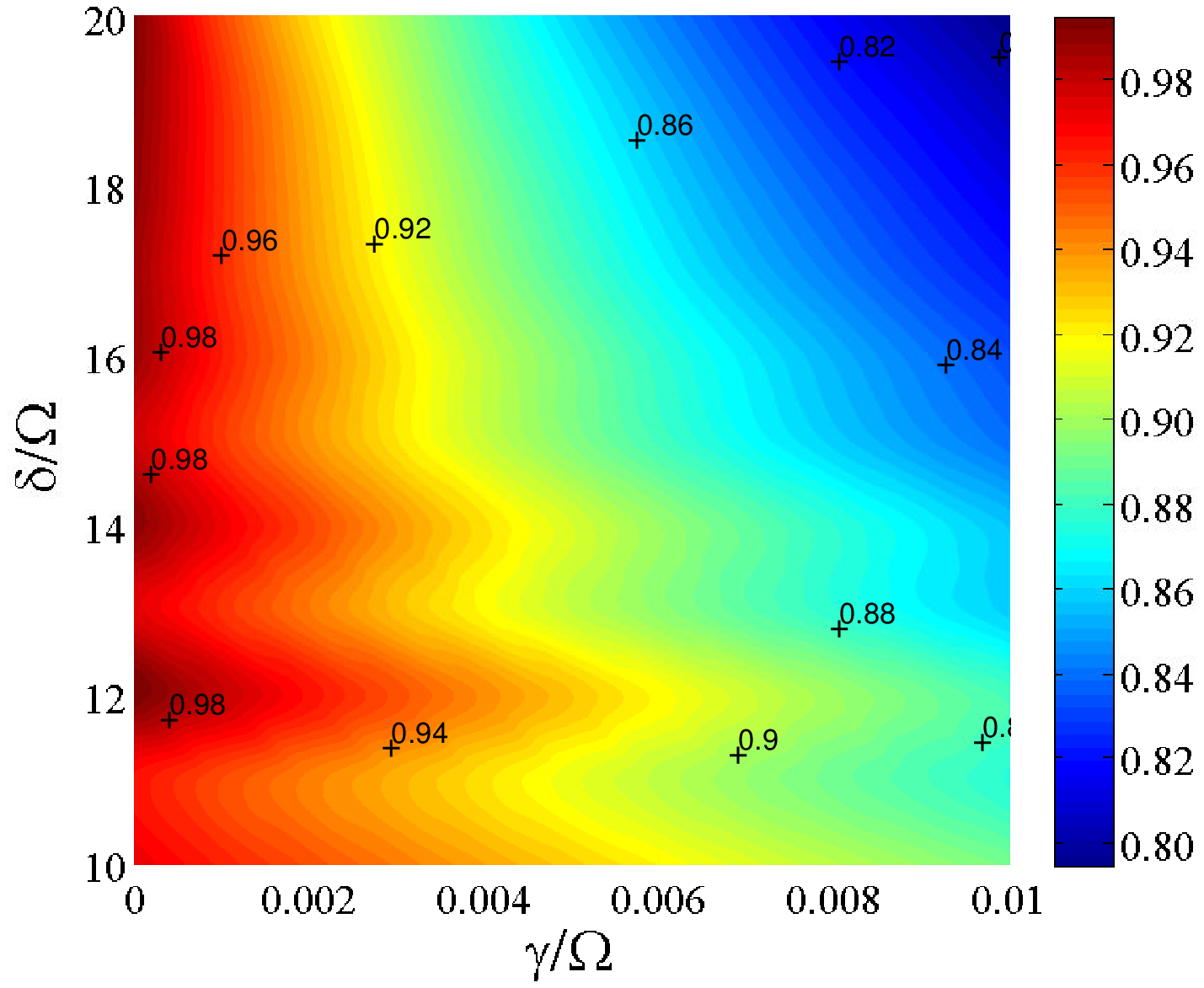} }
\caption{\label{py}(Color online) The fidelity of $|{\rm GHZ}\rangle$ state (left) and controlled-phase gate (right) versus the two-photon detuning $\delta/\Omega$ and decay rate $\gamma/\Omega$ arising from spontaneous emission of Rydberg state $|r\rangle$.}
\end{figure*}

\begin{eqnarray}\label{111}
|\Psi(0)\rangle&=&c_1|000\rangle+c_2|001\rangle+c_3|010\rangle+c_4|011\rangle
\nonumber\\&&+c_5|100\rangle+c_6|101\rangle+c_7|110\rangle+c_8|111\rangle,
\end{eqnarray}
where $c_i$ is the corresponding amplitude of probability obeying the normalization $\sum_i|c_i|^2=1$. The state $|0\rangle$ is decoupled to the dynamical evolution, and
the single-atom state $|1\rangle$ dispersively interacts with the Rydberg state $|r\rangle$ with Rabi frequency $\Omega$, detuned by $-\delta$, while the two-atom state $|11\rangle$ couples to the Bell state $(|1r\rangle+|r1\rangle)/\sqrt{2}$ with Rabi frequency $\sqrt{2}\Omega$, detuned by $-\delta$. Thus the large-detuned dynamical evolution combines
the stark-shift terms of states $|1\rangle_i$ and Rabi oscillation of state $|111\rangle$.
At an arbitrary time $t$, the input state $|\Psi(0)\rangle$ evolves to
\begin{eqnarray}\label{111}
|\Psi(t)\rangle&=&c_1|000\rangle+e^{-i\frac{\Omega^2}{\delta}t}c_2|001\rangle
+e^{-i\frac{\Omega^2}{\delta}t}c_3|010\rangle\nonumber\\&&
+e^{-i\frac{2\Omega^2}{\delta}t}c_4|011\rangle
+e^{-i\frac{\Omega^2}{\delta}t}c_5|100\rangle\nonumber\\&&
+e^{-i\frac{2\Omega^2}{\delta}t}c_6|101\rangle
+e^{-i\frac{2\Omega^2}{\delta}t}c_7|110\rangle\nonumber\\&&
+e^{-i\frac{3\Omega^2}{\delta}t}\cos\big[6\Omega^3t/\delta^2\big]c_8|111\rangle.
\end{eqnarray}
After $|111\rangle$ undergoes a full Rabi oscillation $6\Omega^3t/\delta^2=\pi$, the phase acquired by each computational basis is $\{1,e^{i\alpha},e^{i\alpha}
,e^{2i\alpha},e^{i\alpha},e^{2i\alpha},e^{2i\alpha},-e^{3i\alpha}\}$, where $\alpha=-\delta\pi/(6\Omega)\}$. This transformation does an entangling operation for it is locally unitary equivalent to the stand controlled-phase gate $\{1,1,1
,1,1,1,1,-1\}$, the phase factor $\alpha$ can be canceled by sequentially single-qubit $Z$ gates on $|1\rangle_i$. In the left panel of Fig.~\ref{pzz}, we plot the fidelity of gate by considering an entangling operation on the input state $|\Psi(0)\rangle$ with equivalent weight of computational through the definition $F(t)=\frac{1}{8}|{\rm tr}[U^{\dag}(t)U_{\rm phase}]|$ \cite{stoj}. The different curves correspond to different
values of the two-photon detuning $\delta$. A larger ratio $\delta/\Omega$ naturally creates a higher gate fidelity accompanying a longer interaction time. Among these curves, we find a selection of $\delta=12\Omega$ (black curve) just corresponds to a genuine three-qubit controlled-phase gate because of $\alpha=2\pi$, and the right panel of Fig.~\ref{pzz} characterizes a deviation of $\delta$ for executing such a quantum gate. In the ideal case, the gate fidelity approaches 99.40\%, which is much higher than the scheme proposed by Wu {\it et al.} \cite{wu}, even  when $\delta=11$ or $\delta=13$, a fidelity above $97\%$ is still achievable. In this sense, our scheme is robust against the mismatch of experimental parameters. In addition, although the Rydberg state lifetime   is relative long, it is necessary to assess the effect of spontaneous emission on the performance of entangling and gating operation. To evaluate
the dynamics in the presence of this dissipative effect,
we solve the master equation
$\dot{\hat{\rho}}=i[\hat{H}_I^{'},\hat{\rho}]
+\sum_{i=1}^3\hat{O}^{(j)\dag}_i\hat{\rho}\hat{O}^{(j)}_i,$
where $\hat{O}_i^{(0)}=\sqrt{\gamma_0}|0\rangle_i\langle r|$ and $\hat{O}_i^{(1)}=\sqrt{\gamma_1}|1\rangle_i\langle r|$, and we assume the branching rations of the atom decay from level $|r\rangle$ to $|0\rangle$ and $|1\rangle$ are equal $\gamma_0=\gamma_1=\gamma/2$. Fig.~\ref{py} illustrates the contours of fidelities for
preparation of GHZ state (left) and quantum gate (right) versus the decay rate $\gamma/\Omega$ and two-photon detuning $\delta/\Omega$, respectively. For both cases, the fidelities raise in an oscillating form as increase of $\delta$, and they are away from the optimal value in the presence of $\gamma$. Nevertheless, we can still obtain an
acceptable fidelity under different situations. For example, as $\gamma=0$, $\delta=14\Omega$,
the fidelity for GHZ state is able to reach 99.46\%; and $\delta=10\Omega$ guarantees the fidelity exceeding 90\% at a large decay rate $\gamma=0.01\Omega$. A similar conclusion can be made on the gate operation. Experimentally \cite{steck,japan}, the lifetime of Rydberg state is about $2\pi\times4.8$kHz, the Rabi frequency $\Omega_R$ of the red laser is $2\pi\times25$MHz,
and the Rabi frequency $\Omega_B$ of the blue laser is $2\pi\times300$MHz. To adiabatically eliminate the optical state $|p\rangle$, we choose $\Delta=10\Omega_B$, and these parameters correspond to an effective decay rate $\gamma/\Omega\sim0.002$. By substituting these values into the master equation, we have the fidelities for generation of the entanglement state 96.75\% as $\delta=14\Omega$ and for the controlled-phase gate 96.54\% as $\delta=12\Omega$, which shows that the current scheme is a robust one.

\section{Extension to multi-qubit case}
\begin{figure*}
\scalebox{0.55}{\includegraphics{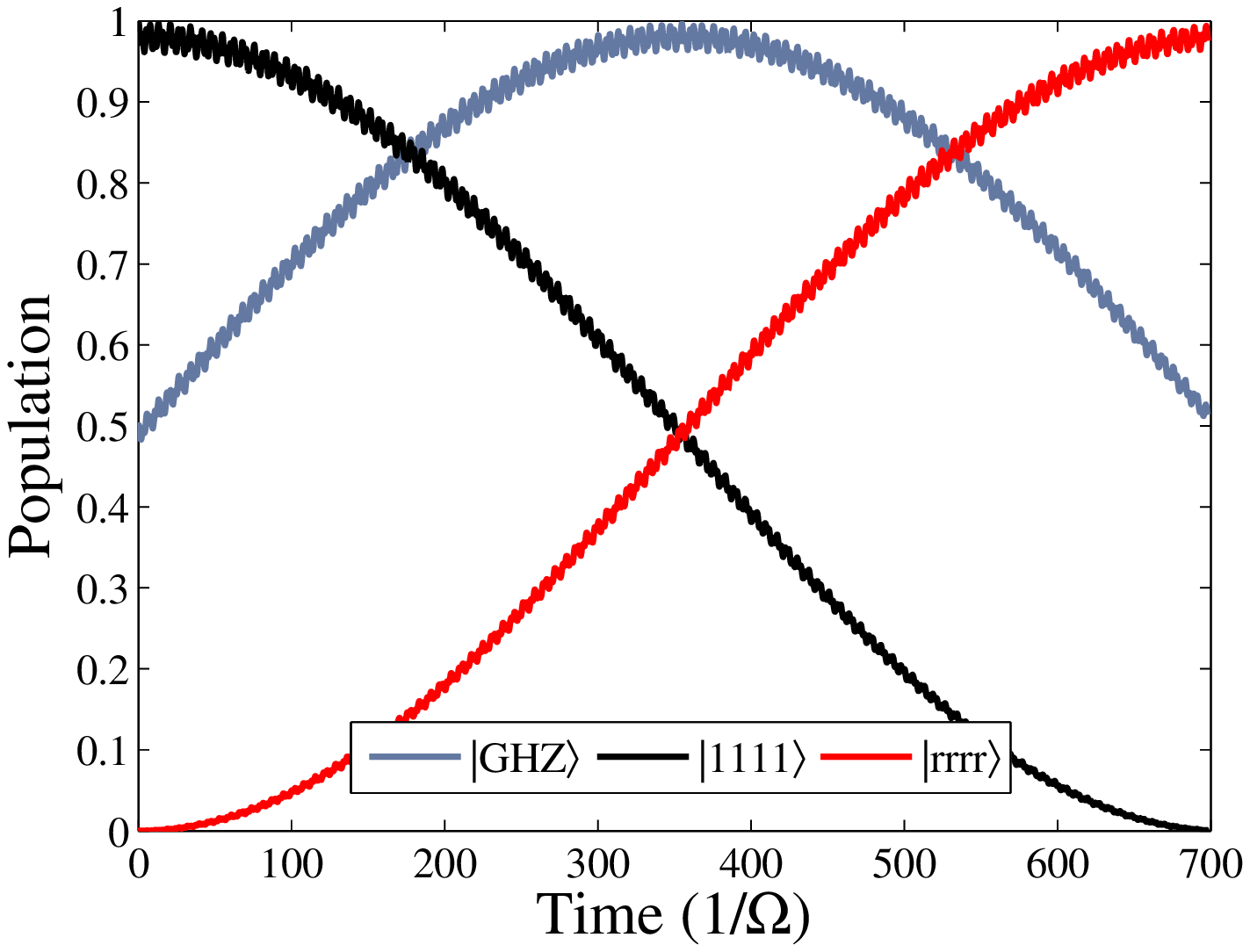}}
\scalebox{0.55}{\includegraphics{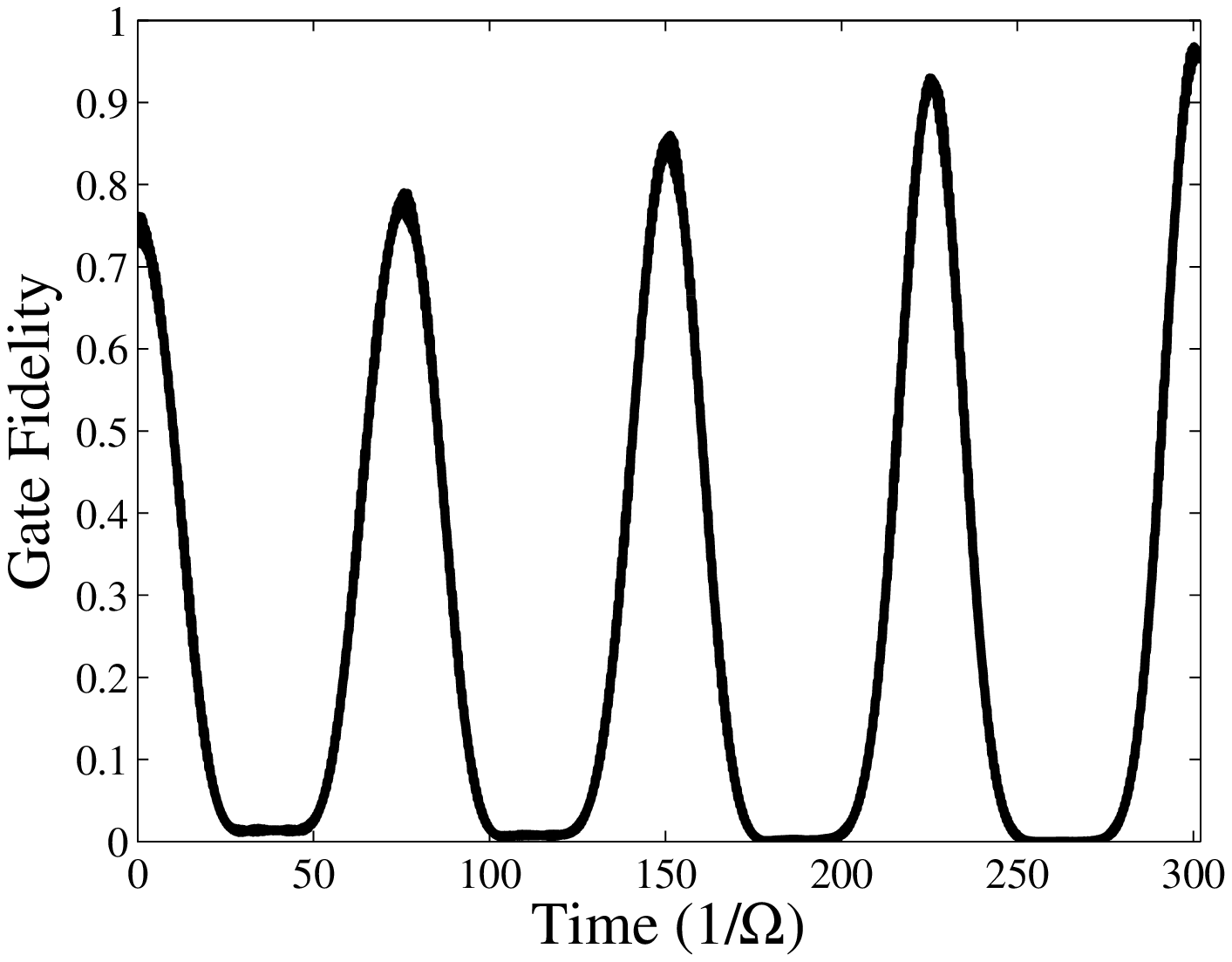} }
\caption{\label{pz}(Color online) Time evolution of fidelities for a four-qubit GHZ state and a controlled-phase gate, where we have set $\delta=20\Omega$ and $\delta=12\Omega$, respectively.  }
\end{figure*}
In general, the dynamical behavior for a quantum system involving many particles is hard to
determined because of the complex interaction among particles. Thus gaining an effective multipartite interaction helpful for quantum information processing even in theory is not only interesting but also meets the requirement of technological development. In this section, we aim to generalize our proposal to achieving a multipartite entangled states and performing a multi-qubit logic gate from the theoretical point of view. Although there may be many limitations that constrain this extension in reality, it will provides a new mechanism on the quantum information processing with neutral atoms. We initialize $N$ atoms in the ground state $|1_1,1_2,1_3,\cdots,1_N\rangle$,
under the action of two-photon transition, this state will interact with an $N$-atom single-excitated state $|N^1\rangle=1/\sqrt{N}(|r_1,1_2,1_3,\cdots,1_N\rangle
+|1_1,r_2,1_3,\cdots,1_N\rangle+|1_1,1_2,r_3,\cdots,1_N\rangle
+\cdots+|1_1,1_2,1_3,\cdots,r_N\rangle)$ with coupling constant $\sqrt{N}\Omega$, detuning $-\delta$. And the state $|N^1\rangle$ further climbs upward a higher collective bi-excited state and so on, till the $N$-excited Rydberg state $|r_1,r_2,r_3,\cdots,r_N\rangle$. The Hamiltonian can then be expanded in the following matrix form
\begin{widetext}
\begin{eqnarray}\label{one}
\hat{H}^{\rm I}_{N}=\left[\begin{array}{c c c c c c c}
0 & \sqrt{C^1_N}\Omega &0 &\cdots & 0 & 0 & 0 \\
\sqrt{C^1_N}\Omega & -\delta & \sqrt{C^2_N}\Omega &\cdots  & 0 & 0 & 0 \\
0 & \sqrt{C^2_N}\Omega & U-2\delta &\cdots & 0  & 0 & 0 \\
\vdots & \vdots & \vdots & \ddots & \vdots & \vdots & \vdots \\
0 & 0 & 0  &\cdots  & C^2_{N-2}U-(N-2)\delta & \sqrt{C^{N-2}_N}\Omega & 0  \\
0 & 0 & 0  &\cdots  & \sqrt{C^{N-2}_N}\Omega & C^2_{N-1}U-(N-1)\delta& \sqrt{C^{N-1}_N}\Omega \\
0 & 0 & 0  &\cdots  & 0 & \sqrt{C^{N-1}_N}\Omega & C^2_{N}U-N\delta  \\
\end{array}
\right],
\end{eqnarray}
\end{widetext}
where $\sqrt{C^{i}_N}$ denotes the binomial coefficient. For the purpose of directly
coupling $|1_1,1_2,1_3,\cdots,1_N\rangle$ and $|r_1,r_2,r_3,\cdots,r_N\rangle$, we
regulate $C^2_{N}U-N\delta=0$, then the Hamiltonian can be written as
\begin{eqnarray}\label{111}
\hat{H}^{\rm I}_{N}&=&\sum^{N-1}_{i=1}\sqrt{C_N^i}\Omega|N^i\rangle\langle N^{i+1}|+{\rm H.c.}\nonumber\\&&+\sum_{j=1}^N
\frac{(j-N)j}{N-1}\delta|N^j\rangle\langle N^j|.
\end{eqnarray}
It is easy to check that both the diagonal and the skew are mirror symmetry, so we can have a Raman coupling between the ground state and Rydber state in virtue of $N$-order perturbation theory. To test and verify our assumption, we numerically simulate the four-qubit GHZ state and quantum gate in Fig.~\ref{pz}, where we have set $\delta/\Omega=20$ to get a good approximation for entangled preparation and $\delta/\Omega=12$ to achieve a stand four-qubit controlled-phase gate.
\section{Conclusion}

In conclusion, we have presented a robust way for one-step generation of tripartite
GHZ state and controlled-phase gate on neutral Rydberg atoms. This scheme works well in the regime where the Rydberg interaction holds a comparable strength to the two-photon detuning. An $N$-qubit GHZ state and quantum logic gate are then straightly generalized, which can be confirmed by the simulation a four-particle system. Our method provides an alternative mechanism for creation of multipartite GHZ state, and we believe our work will be useful for the experimental realization of quantum
information with neutral atoms in the near future.
\begin{center}{\bf{ACKNOWLEDGMENT}}
\end{center}

This work is supported by Fundamental Research Funds for the Central Universities under Grant No. 12SSXM001,  National
Natural Science Foundation of China under Grant Nos. 11204028 and 11175044, and National Research
Foundation and Ministry of Education, Singapore (Grant
No. WBS: R-710-000-008-271). X. Q. Shao was also supported in part by the Government of China through CSC.
\

\end{document}